# Broadband high-resolution molecular spectroscopy with interleaved mid-infrared frequency combs

A. V. Muraviev, D. Konnov, and K. L. Vodopyanov
CREOL, College of Optics and Photonics, Univ. Cent. Florida, Orlando, FL 32816, USA
* vodopyanov@creol.ucf.edu

**Historically, there has been a trade-off in spectroscopic measurements between high spectral resolution, broadband coverage, and acquisition time. Optical frequency combs, initially envisioned for precision spectroscopy of the hydrogen atom in the ultraviolet region, are now commonly used for probing molecular rotational-vibrational transitions throughout broad spectral bands in the mid-infrared with superior accuracy, resolution and speed. Here we demonstrate acquisition of 2.5 million spectral points over the continuous wavelength range of 3.2–5.1 μm (frequency span 1200 cm$^{-1}$, resolution ~100 kHz) via interleaving comb-tooth-resolved spectra acquired with a highly-coherent broadband dual-frequency-comb system based on optical subharmonic generation. With the original comb-line spacing of 115 MHz, overlaying spectra with gradually shifted comb lines we fully resolved amplitude and phase spectra of molecules with narrow Doppler-broadened lines, such as $v_1+v_3$ band of carbon disulfide ($CS_2$) and its isotopologues. Also, we were able to observe the chemical oxidation reaction of $CS_2$ by detecting traces of carbonyl sulfide (OCS) via its asymmetric $v_3$ stretch band.**

## Introduction

Coherent laser beams in the 3 to 20 μm mid-infrared (mid-IR) region provide a unique prospect for sensing molecules through their strongest absorption bands. Thanks to their coherent and broadband nature, optical frequency combs can probe molecular signatures over an extensive spectral span simultaneously [1]. When tightly phase locked to a narrow-linewidth reference laser, a frequency comb is equivalent to an enormous number ($10^5$–$10^6$) of narrow-band optical oscillators evenly spaced across the spectrum, each with a frequency accuracy and linewidth equal to that of the reference laser [2]. In terms of the highest spectral resolution, frequency comb spectroscopy is on par with tunable laser spectroscopy, however the former has the advantage of massive parallelism of data collection and, most importantly, the absolute optical frequency referencing to an external standard (e.g. atomic clock) over the whole bandwidth of the comb.

Laser combs with high degree of phase coherence can be used in dual comb spectroscopy (DCS) – one of the most advanced techniques that is well suited for getting spectroscopic data with high precision and resolution. In a dual-comb spectrometer, a sensing comb is transmitted through a sample and then multi-heterodyned against a local oscillator (LO) comb which has a repetition rate $f_r$ that differs by a small fraction $\Delta f_r$ from that of the sensing comb. Compared to classical Fourier transform infrared spectroscopy, DCS demonstrates remarkable improvement of spectral resolution, data acquisition speed, and sensitivity, all at the same time [2].

With high degree of mutual coherence between the two combs in a DCS system, it is possible to obtain comb-tooth resolved spectra [2,3,4]. Using DCS in the near-IR range with a comb spanning 1360–1690 nm, Zolot et al. resolved the phase and amplitude of over 400,000 individual modes at a mode spacing of 100 MHz [4]. In the mid-IR, by utilizing combs near 3.3 µm with a spectral span of 12 cm$^{-1}$, achieved by down converting near-IR electro-optic modulation (EOM) combs via difference-frequency generation, Yan et al. resolved 1200 comb lines with a line spacing of 300 MHz [5]. Subsequently, two groups demonstrated mode resolved spectra over a very broad range of frequencies: Ycas et al. resolved 270,000 comb lines between 2.6 and 5.2 µm in four overlapping sequences (mode spacing 200 MHz, combined frequency span 1800 cm$^{-1}$) [6], and Muraviev et al. resolved 350,000 comb lines with a finesse of 4,000, within a single comb spanning 3.1–5.5-µm (mode spacing 115 MHz, frequency span 1400 cm$^{-1}$) [7]. In the same work, the authors demonstrated simultaneous detection of more than 20 molecular species in a mixture of gases.

Once comb teeth are resolved, the spectral resolution is defined by an absolute comb tooth linewidth, which can be orders of magnitude narrower than the intermodal spacing; high-resolution measurements then can be implemented by interleaving spectra taken with discretely stepped either comb repetition rate $f_r$ or carrier-envelope offset (CEO) frequency $f_{ceo}$ [2,8,9,10,11,12]. Baumann et al. used mid-IR difference frequency combs near 3.4 µm with spectral span 30 cm$^{-1}$, to attain a high-resolution spectrum of methane, by interleaving spectra acquired by shifting the combs by 25 MHz intervals – one-quarter of the comb spectral spacing of 100 MHz [8]. Using quantum cascade laser (QCL) combs, Villares et al. performed DCS measurements in the 7-µm region (span 16 cm$^{-1}$) with the spectral resolution that was improved from the original comb spacing of 7.5 GHz to 80 MHz by frequency sweeping the combs via current modulation [13]. Similarly, by utilizing dual QCL combs near 8.3 µm (span 55 cm$^{-1}$), Gianella et al. achieved an improvement in the spectral resolution from 9.8-GHz to 30 MHz, by sweeping the frequencies of both the sensing and the local combs via synchronized current modulation [14]. In the two above QCL scenarios the combs were free running with no absolute referencing of the optical frequency; rather, they were calibrated via comparing the spectra with the HITRAN database [15]. Overall, the spectral coverage of mid-IR measurements with interleaved combs does not exceed 60 cm$^{-1}$ with one exception of a silicon microresonator comb with a span of 3–3.5 µm that was scanned with a step of 80 MHz via tuning both the pump laser and the cavity resonance; however the comb lines were scanned over only 16 GHz – a small portion of the 127-GHz mode spacing [16].

One of the challenges in DCS system development – for example in applications related to multi-species spectral analysis in gas mixtures – is the requirement for both broad spectral coverage and high spectral resolution. Such a combination requires high mutual coherence between sensing and LO combs [2].

Degenerate (subharmonic) optical parametric oscillators (OPOs) pumped by mode-locked lasers are noteworthy sources of broadband mid-IR frequency combs [17,18,19,20,21] and are now used in spectroscopic studies [7,22], random number generation [23], and in coherent Ising machines [24]. Their key benefits are: low (~10 mW) oscillation threshold, extremely broad bandwidth, good stability when the cavity is actively locked to resonance, and high conversion efficiency that can exceed 50% [25]. It has been established that a subharmonic OPO is an ideal coherent frequency divider without any excess phase noise, which rigorously both coherently down-converts and augments the spectrum of the pump frequency comb [26, 27, 28, 29].

Here, using our highly-coherent DCS system based on subharmonic OPOs, we demonstrate the acquisition of 2.5 million spectral data points over the whole (no gaps) spectrum of 3.2–5.1 µm (span 1200 cm$^{-1}$). The spectral sampling achieved by interleaving spectra obtained with consecutively shifted combs was adequate to fully resolve the whole Doppler-broadened absorption band $\nu_1+\nu_3$ of carbon disulfide ($CS_2$) at 4.5–4.7 µm and generate amplitude and phase spectra for several other molecules (CO, OCS) that are referenced to an accurate frequency standard.

## Results

**Experimental setup**. Our dual-comb system (see Methods) used a pair of subharmonic OPOs based on orientation-patterned GaAs (OP-GaAs) crystal as a $\chi^{(2)}$ gain medium, pumped by a highly coherent twin Tm-fiber frequency comb system with a central wavelength of 1.93 µm, pulse duration of 90 fs, repetition frequency of 115 MHz, and the average power of 300 mW for each laser [7].

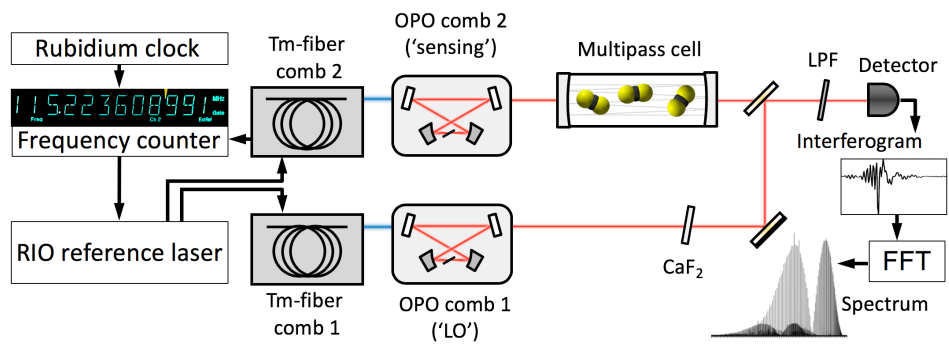

Fig. 1 Schematic of the dual-comb spectroscopy setup. LPF, longwave pass filter ($\lambda> 3$ µm); $CaF_2$, a dielectric plate to compensate the dispersion of the multipass cell windows.

In order to stabilize Tm laser frequency combs, a portion of each laser's output was used to generate a supercontinuum (SC) in a nonlinear silica fiber. While the 1.1-µm and 2.2-µm SC components were used to stabilize the CEO frequency via $f$–$2f$ interferometry, the component near 1.56 µm was utilized to obtain beat notes with a narrow-linewidth reference diode laser from Redfern Integrated Optics (RIO). Due to the tight phase locking to the common optical reference, the comb teeth linewidths for both Tm-fiber lasers were similar to that of the RIO laser (~3 kHz) [27]. Each OPO (Fig. 1) had a ring-cavity bow-tie design with low cavity group delay dispersion (GDD) achieved by (i) using low-dispersion mirrors, (ii) a thin (0.5-mm-long) OP-GaAs crystal, and (iii) an intracavity wedge made of $CaF_2$ for GDD compensation [7,28]. The instantaneous spectral coverage of the DCS system was 3.17–5.13 µm (span 1200 cm$^{-1}$) at -10 dB and 3.08–5.40 µm (span 1400 cm$^{-1}$) at -20 dB level. The mutual coherence time between the two subharmonic OPOs was previously measured to be as long as ~ 40 s [7]. Our data acquisition electronics and frequency counters were referenced to a Rb atomic clock, which provided $10^{-9}$–$10^{-10}$ absolute accuracy of the frequency readings.

**Spectroscopy with interleaved combs**
We used a one-sided DCS configuration (Fig.1), where only one ('sensing') comb passes through a molecular sample. This allows to measure the phase spectrum and eliminates the spectral data split inherent to the double-sided scheme, where the absorption profile is probed by two comb lines simultaneously. For absorption measurements we used a multipass gas cell (AMAC-76LW from Aerodyne) with 76-m path length and 0.5-liter volume. The sensing and

LO combs were combined after the gas cell and sent to an infrared detector (InSb from Kolmar, 77 K, 60 MHz, longwave cut-off 5.6 µm), whose output was fed into a 16-bit analogue-to-digital converter (AlazarTech, ATS9626). In order to make the shape of the interferogram more favourable for triggering the data acquisition process, we used a $CaF_2$ dielectric plate (Fig. 1) to compensate the dispersion caused by the windows of the multipass cell.

The main focus of this study was carbon disulfide ($CS_2$) molecule. The choice of $CS_2$ was motivated by its importance for atmospheric chemistry [30], astrobiology [31], and medical diagnostics [32,33], while accurate high-resolution spectroscopic data for $CS_2$ in the mid-IR range are not available. Also, its high molecular weight results in narrow Doppler-broadened absorption lines at room temperature and allows us to demonstrate the potential of our DCS system.

A liquid sample of $CS_2$ dissolved in n-hexane at a concertation of 100 µg/ml from Chem Services, Inc. was mixed with ambient air and evaporated under vacuum condition into the gas cell. A small amount of CO gas was added to the mixture for the frequency scale accuracy verification purposes. The gas mixture in the cell had a total pressure of 6.03 mbar and contained $CS_2$ at a concentration of 94.6 ppm (part-per-million by volume), CO at 7 ppm, and trace gases present in the room air ($CO_2$, $H_2O$, and $N_2O$). According to the HITRAN database, the absorption spectrum of n-hexane in our spectral range does not have any sharp features– its absorption resulted only in an offset baseline, which was subtracted from the spectral data. Due to low pressure of the gas mixture and high spectral resolution, the absorption lines of different species were well separated and practically didn't interfere with each other.

First, we demonstrated that an extensive amount of spectral information could be obtained by interleaving comb-line-resolved DCS spectra over the whole comb span (3.2–5.1 µm). The repetition rate offset between the two combs of $\Delta f_r$=138.5 Hz, allowed us to map the whole 1200 $cm^{-1}$-wide optical spectrum into a radiofrequency (RF) range of 1–48 MHz ($<f_r/2$, see Methods). For each comb-line-resolved measurement, we coherently averaged data streams consisting of 10 centerbursts spaced by a $1/\Delta f_r$=7.2 ms, with the coherent averaging time from a few seconds to several hours. For each data stream consisting of ~10 M data points, the data acquisition process was triggered by a sharp central spike of the interferogram, providing good repeatability of the waveforms. This allowed long coherent averaging without any software phase correction algorithms.

After Fourier transforming the time-domain signal and RF-to-optical frequency up-scaling (see Methods) we obtained a comb-tooth-resolved optical spectrum. By changing the frequency of the reference RIO laser, the comb lines were tuned, thus filling the 115-MHz gap between the comb lines. With an estimated comb-tooth linewidth for our OPOs of ~ 20 kHz [7], in theory, one can achieve by interleaving the spectra, more than $10^9$ spectral points of resolution. In our case, to fully resolve Doppler-broadened linewidths of species with high molecular weight, such as $CS_2$ (Doppler linewidth 90 MHz), we chose to interleave eight comb-line-resolved spectra. The spectra were acquired by step-scanning the RIO laser frequency in approximately 42-MHz increments, resulting (at a fixed $f_{ceo}$) in stretching the entire frequency combs and providing 14.4 MHz comb-line increments, ~ 1/8 of the original 115-MHz comb-line spacing, in the vicinity of $CS_2$ $\nu_1+\nu_3$ absorption band at 2120-2200 $cm^{-1}$.

Fig. 2a shows the comb-tooth-resolved optical spectrum spanning 3.17–5.13 µm, consisting of 2.5 million interleaved comb lines. The total averaging time was 200 min (with 25 min for each frequency shifted comb). The expanded spectrum of Fig. 2b reveals absorption dips due to the molecules present in the gas cell ($^{13}CO_2$, $N_2O$, $CS_2$, CO, OCS, and $H_2O$). With a further x-scale expansion (Fig. 2c) one can see interleaved comb lines modulated by a $CS_2$ absorption feature.

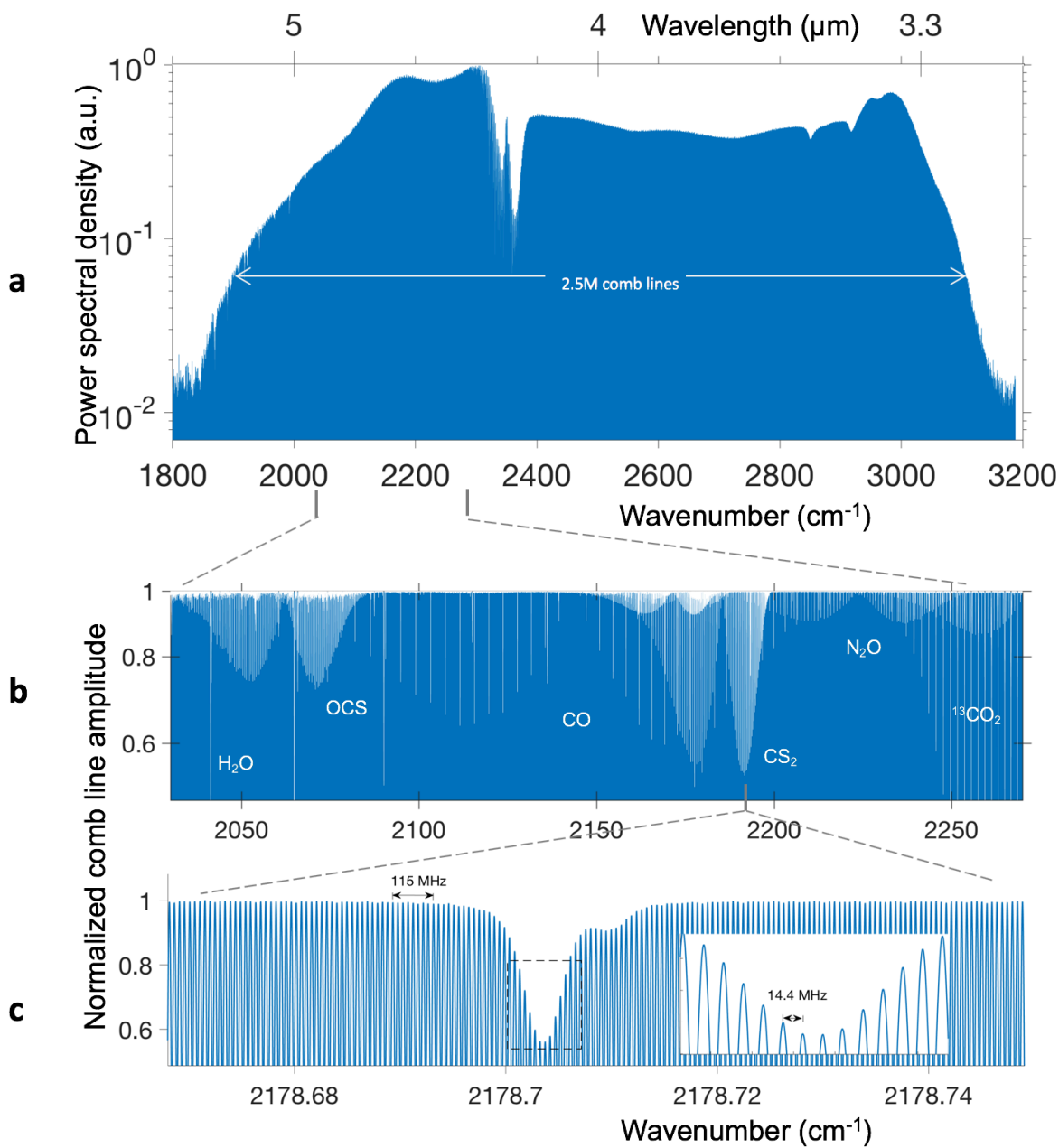

Fig. 2. Interleaved comb-tooth-resolved optical spectrum. **a**, Spectrum containing 2.5 million comb lines, obtained by 8 interleaved measurements. **b**, Expanded and normalized spectrum showing dips due to molecular absorption. **c**, Further zooming shows individual interleaved comb lines and a dip due to a $CS_2$ absorption feature.

For the measurements focused on $CS_2$, we narrowed down the spectral span to 2000-2400 cm$^{-1}$ (4.2–5.0 μm) by using an optical bandpass filter. This increased SNR (which scales, at a fixed power at the detector, as the inverse of the total number of comb lines [9]) and allowed 4 times increase of $\Delta f_r$ to 554 Hz, which accelerated, by the same amount, the data acquisition rate. Figs. 3a,b show the amplitude and phase spectra from eight interleaved DCS measurements, corresponding to the $v_1+v_3$ absorption band of $CS_2$, with the acquisition time for each comb-line resolved spectrum of 200 min. The spectrum (Figs. 3a) also contains absorption lines of CO and $N_2O$ molecules, present in the gas mixture. Each data point in Fig. 3 represents a fitted amplitude of a comb line. The distance between filled circles in the expanded view (Fig. 3c) correspond to the original comb-line spacing (115 MHz), while the interleaved data points (open circles) are spaced by about 1/8 of that value (14.4 MHz) in this spectral region.

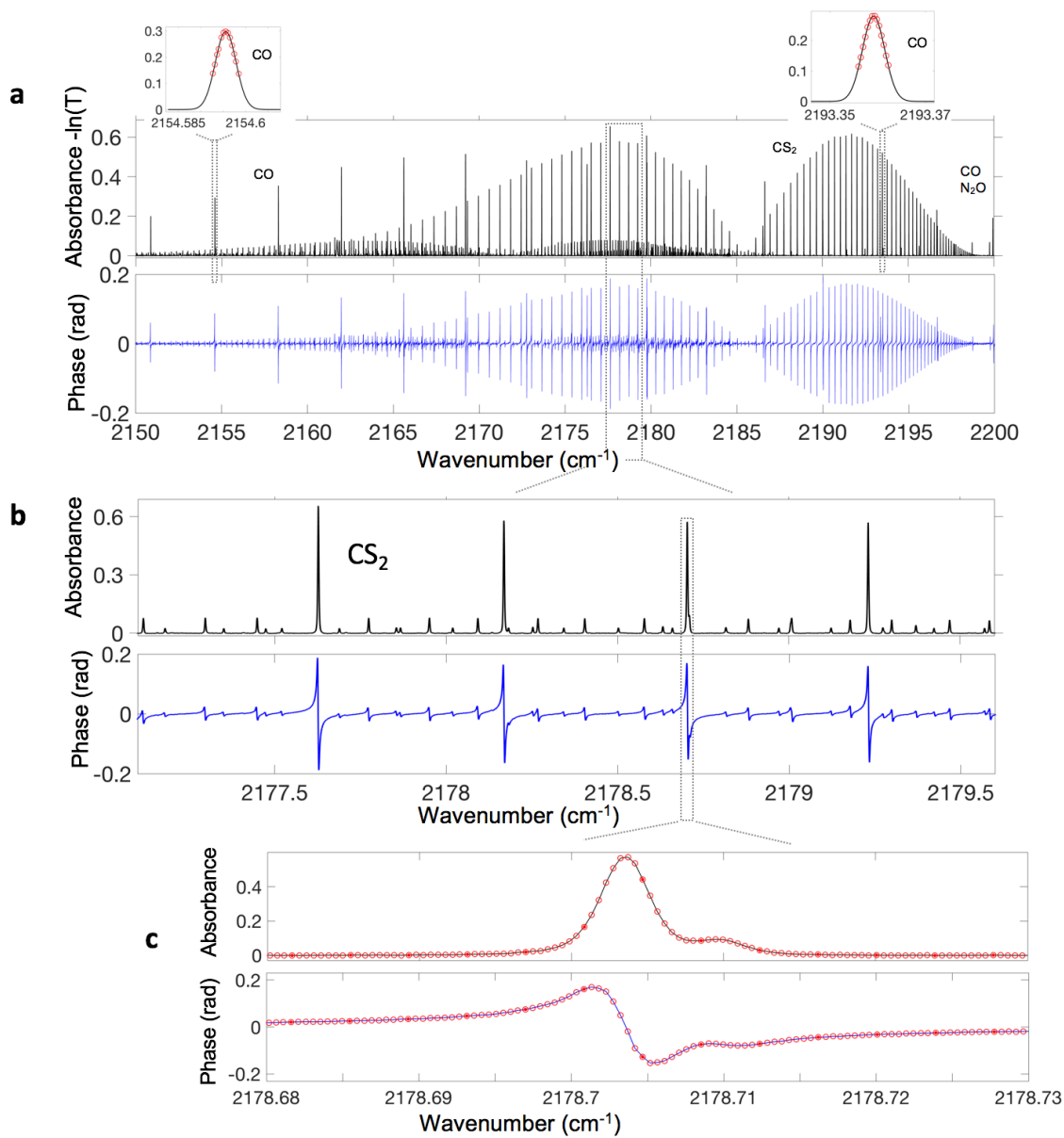

Fig. 3. $CS_2$ $v_1+v_3$ rovibrational spectrum at 14 MHz point spacing. **a**, Amplitude and phase spectrum, obtained from interleaved DCS measurements, corresponding to the $v_1+v_3$ absorption band of $CS_2$, plus absorption features of CO and $N_2O$ present in the cell. The insets on the top show the measured (circles) and the HITRAN database (solid line) peaks of CO, demonstrating high absolute accuracy of the frequency scale. **b**, Expanded $CS_2$ amplitude and phase spectrum. **c**, $CS_2$ spectrum with further expanded x-scale. Circles correspond to interleaved DCS measurements and solid lines – to the fitted curves. The distance between the filled circles correspond to the original comb-line spacing (115 MHz).

Since the CO spectral lines are well characterized in HITRAN, the presence of CO allowed us to verify high precision of the absolute frequency referencing of our measurements. The insets in Fig. 3a show measured (circles) and simulated (HITRAN, solid lines) absorption peaks of CO near 2155 and 2193 cm$^{-1}$; the measured positions of the peaks' maxima deviate from those obtained from HITRAN by less than 1 MHz.

Fig. 4 shows the log-scale absorbance spectrum of $CS_2$ with the weaker bands visible. A high resolution and long coherent averaging allowed us to identify 981 spectral lines of the $\nu_1+\nu_3$ absorption band of $CS_2$ with the signal-to-noise ratio (SNR) of $3\times10^3$. In addition to the main molecule, we were able to resolve the spectra of three isotopologues ($^{34}S^{12}C^{32}S$, $^{33}S^{12}C^{32}S$, and $^{32}S^{13}C^{32}S$), as well the spectra of seven hot bands due to transitions from the thermally excited vibrational states [34]; some of these bands are labelled in Fig.4. The inset shows the two peaks corresponding to the low-abundance $^{33}S$ isotope, along with the two peaks for the main molecule.

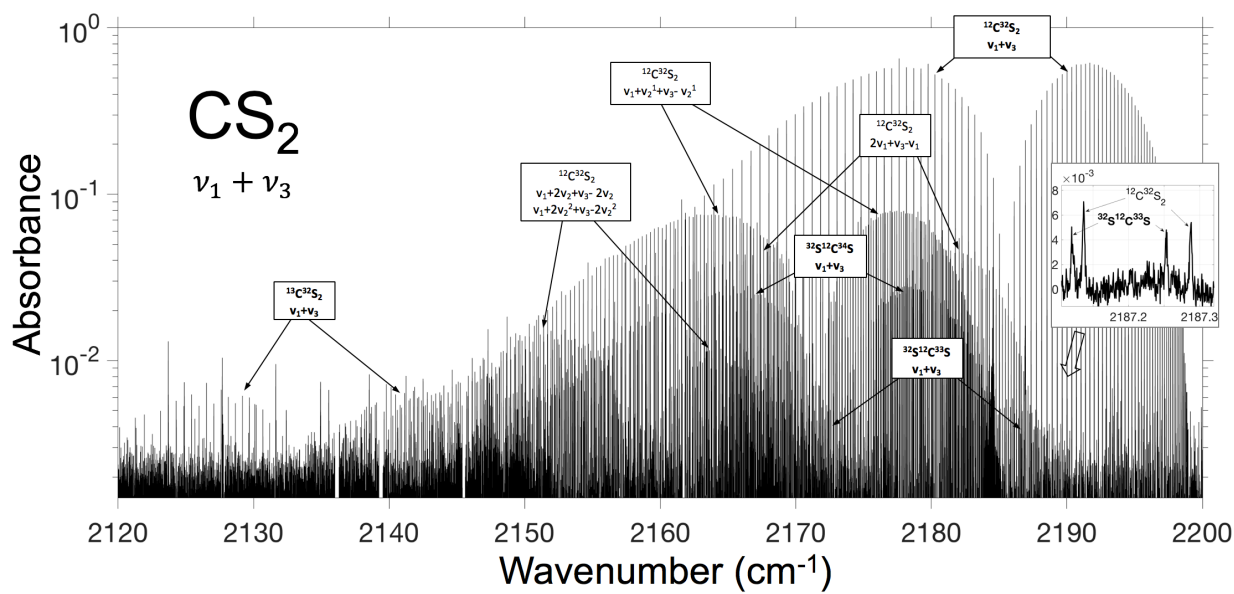

Fig. 4. Log-scale $CS_2$ spectrum revealing isotopologues and hot bands. The absorption bands include: (i) bands due to isotopes: $^{34}S$, $^{33}S$, and $^{13}C$, (ii) hot bands from the thermally excited vibrational states $\nu_1$ and $\nu_2$. The inset shows peaks due to the weak $^{33}S$ isotope, along with those for the main molecule.

Fig. 5 displays the portion of the spectrum (for amplitude and phase) that reveals the presence of trace amounts of carbonyl sulfide (OCS) molecule, although it was not originally present in the gas mixture. The plot also shows the simulated absorbance spectrum – asymmetric $\nu_3$ stretch band of OCS based on HITRAN. Our repeated measurements revealed that the OCS concertation (here ~ 1.75 ppm) kept rising at approximately 15% per day, while the intensity of $CS_2$ absorption was slightly decreasing, which can be explained by the fact that OCS may be the result of the chemical oxidation reaction of $CS_2$ in the presence of OH and $O_2$ in the gas cell [35].

With the averaging time of 200 min ($\tau$=12,000 s) for a comb-line resolved spectrum in the 2000–2400 cm$^{-1}$ band of and SNR of $3 \times 10^3$, the DCS figure of merit, identified in [9] as $FOM = SNR \times M/\sqrt{\tau}$ ($M$ is the number of comb lines), is equal to $FOM= 2.85 \times 10^6$ Hz$^{1/2}$.

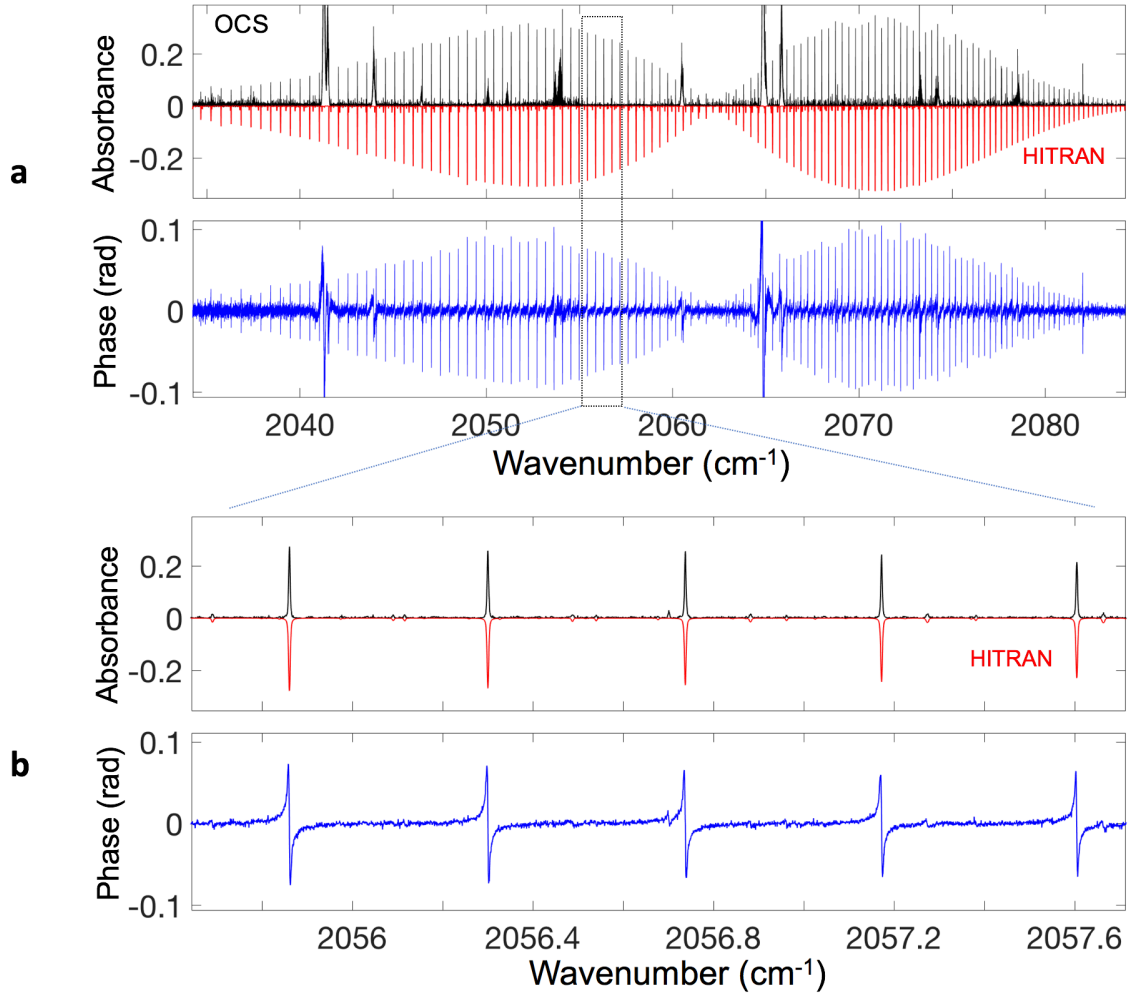

Fig. 5. OCS spectrum. **a**, Amplitude and phase spectrum for the $\nu_3$ asymmetric band of OCS. An extra noise (e.g. at 2041 and 2065 cm$^{-1}$ etc.) is related to strong (and spectrally broadened) water absorption lines outside the gas cell. **b**, Expanded OCS spectrum. The simulated amplitude (absorbance) spectrum based on HITRAN is shown on both plots and is inverted for clarity.

## Conclusion

Using highly coherent broadband dual-comb system based on subharmonic generation from the 1.93-µm pump, we demonstrate acquisition of 2.5 million spectral elements over the wavelength range of 3.2–5.1 µm (frequency span 1200 cm$^{-1}$) achieved by interleaving comb-line resolved spectra with discretely stepped comb-line spacing $f_r$. We measured Doppler-broadened spectra of several molecules with fully resolved amplitude and phase, and, for the first time to our knowledge, characterized the whole $\nu_1+\nu_3$ band (4.5-4.7 µm) of $CS_2$ and its three isotopologues, with high SNR and ~100-kHz optical frequency scale accuracy, thanks to the direct linking to a primary frequency standard. A plethora of generated data can be used as a testing ground for theoretical models, as well as a source for molecular spectroscopic databases, such as HITRAN.

We acknowledge support from the Office of Naval Research (ONR), grant number N00014-15-1-2659 and from the Defense Advanced Research Projects Agency (DARPA), grant number W31P4Q-15-1-0008. We thank Ekaterina Karlovets and Iouli Gordon for assisting with interpretation of the $CS_2$ spectra, and Pete Schunemann for providing nonlinear OP-GaAs crystals.

## Methods

**OPO comb modes and RF-to-optical frequency mapping**. The comb modes for both Tm-fiber pump lasers were locked, via octave-wide supercontinuum generation, at the two common anchor points (Fig. 1S): near zero frequency (via $f$-to-$2f$ beat note) with a carrier-envelope offset (CEO) of $f_{ceo}^{pump}$=+190 MHz (point A), and near the frequency of a narrow-linewidth (~3 kHz) continuous wave (CW) reference 'RIO' laser at λ≈1564 nm with an offset $f_{RIO}$=+140 MHz (point B). Between A and B, there are $N_1$ intermodal intervals ($f_{r1}$) for laser 1 and $N_2$ intervals ($f_{r2}$) for laser 2, such that $N_1 \times f_{r1} = N_2 \times f_{r2}$ and

$$\Delta f_r = f_{r2} - f_{r1} = \frac{N_1-N_2}{N_2} f_{r1} = \frac{\Delta N}{N_2} f_{r1} = \frac{\Delta N}{N_1} f_{r2}. \tag{1}$$

The repetition rate offset $\Delta f_r$ between the two lasers (assume $f_{r2} > f_{r1}$) is quantized, because $\Delta N = N_1 - N_2$ is an integer. For example, for $\Delta N$=2, $\Delta f_r \approx$ 138.5 Hz one consequence of (1) is that for even $N_2$=$2k_0$ and $N_1$=$2k_0+2$ ($k_0$ is an integer), the ratios $f_{r1}/\Delta f_r$ and $f_{r2}/\Delta f_r$ are integer numbers. In fact, in our experiments the measured ratio $f_{r1}/\Delta f_{rep}$ ($\approx k_0$) is an integer: 831844 with a remainder after division < $10^{-3}$. Hence the absolute frequency of the anchor point B gets known once $f_r$ and $\Delta f_r$ are measured (prior knowledge of the frequency of the reference laser is not needed).

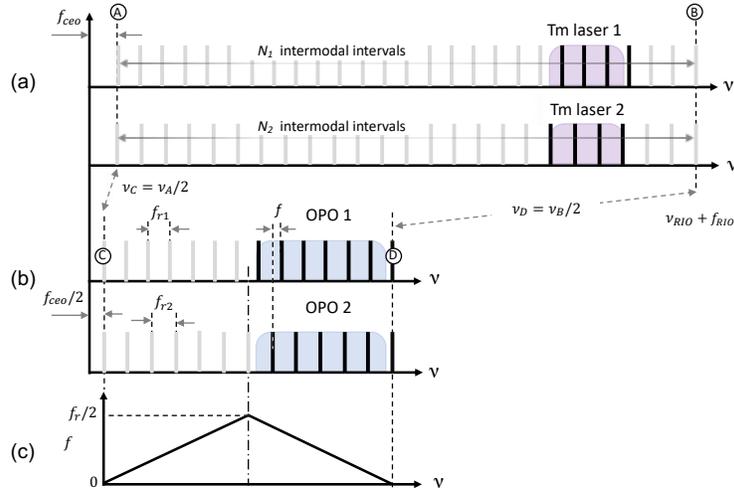

**Fig. 1S** Frequency comb lines. (a) Comb lines (extrapolated to zero frequency) for the Tm pump lasers with repetition frequencies $f_{r1}$ and $f_{r2}$ respectively. The combs are mutually phase-locked, and their lines overlap at the two lock points 'A' and 'B'. (b) Comb lines for the two subharmonic OPOs. The combs are frequency locked to the pump; their teeth overlap at the anchor point 'C' distanced by $f_{ceo}/2$ from zero, and (if $N_1$ and $N_2$ are even) at the anchor point 'D' (λ≈3.13 µm). (c) Illustration of the RF-to-optical frequency mapping for even $N_1$ and $N_2$ and $\Delta N$=2.

Both $OPO_1$ (LO comb) and $OPO_2$ (sensing comb) were running in the frequency-divide-by-2 mode with respect to the pump, such that that CEO frequency for each of them (anchor point C in Fig. 1S) was half of that of the pump (CEO frequencies for both Tm lasers were kept equal)

$$f_{ceo}^{OPO1} = f_{ceo}^{OPO2} = \frac{1}{2} f_{ceo}^{pump}. \tag{2}$$

Another anchor point (D) for the OPOs, with $\nu_D=½\ \nu_B$ (near ~3.13 μm, about twice the wavelength of the CW reference), just above (in frequency) our spectral range of interest (3.2-5.1 μm). Therefore, we used an RF-to-optical mapping (Fig. 1S) for the 'sensing' comb:

$$\nu = \nu_D - \frac{f_{r2}}{\Delta f_r} f = \frac{1}{2} f_{ceo}^{pump} + \frac{f_{r2}}{\Delta f_r}(f_{r1} - f), \tag{3}$$

where $\nu$ is the optical frequency that interrogates the sample, and $f$ is the RF frequency. The whole 1200 cm$^{-1}$ spectral span was mapped to the 1–48 MHz RF range. In a similar way, we performed mapping for larger $\Delta f_r$, e.g. corresponding to $\Delta N=8$, when smaller optical bands were used.

**Active stabilization of the RIO laser**
Since the two Tm-fiber combs are phase locked to a common optical reference (RIO laser), there is a high mutual coherence between these combs, which is transferred to high mutual coherence between the two subharmonic OPOs. Because the frequency of the RIO laser ($\nu_{RIO}$) drifts by >10 MHz during the day, which affects the absolute position of the comb teeth, we implemented its active stabilization. Using frequency counter readings for the repetition rate $f_r$ for one of the Tm-fiber combs as a feedback, and by acting on the RIO laser temperature (the servo loop time constant ~ 4 s), we achieved the standard deviation of the repetition frequency of $\sigma f_r=0.09$ Hz ($\sigma f_r/f_r = 7.8 \times 10^{-10}$), which corresponds to the RIO laser frequency uncertainty $\sigma\nu_{RIO}=140$ kHz. This translates (at a fixed $f_{ceo}$) to the uncertainty of the absolute frequency knowledge (as well as the spectral resolution) of 50 kHz for our comb lines near 4.6-μm CS$_2$ absorption band. Remarkably, with the fixed $\Delta N$, $f_{ceo}^{pump}$ and $f_{RIO}$ frequency offsets, and with just two readings from the frequency counters ($f_{r1}$ and $\Delta f_r = f_{r2} - f_{r1}$), we were able to determine the vacuum wavelength of the reference laser ($\lambda_{RIO} \approx 1563.88$ nm) with the relative accuracy of 10$^{-9}$.